\documentclass[a4paper,fleqn,usenatbib]{mnras}
\usepackage[T1]{fontenc}
\usepackage{ae,aecompl}
\usepackage{graphicx}
\usepackage{amsmath}
\usepackage{amssymb}
\usepackage{epstopdf}

\newcommand{\hii}    {H\,{\sc{ii}}~}
\defcitealias{paper_mnrasi}{Paper~I}

\title[Dust dynamics and evolution in \hii regions]
{Dust dynamics and evolution in \hii regions. II. Effects of dynamical coupling between dust and gas}

\author[V. V. Akimkin, M. S. Kirsanova, Ya. N. Pavlyuchenkov, D. S. Wiebe]{
V. V. Akimkin, M. S. Kirsanova, Ya. N. Pavlyuchenkov\thanks{E-mail: pavyar@inasan.ru}, D. S. Wiebe
\\
Institute of Astronomy of the Russian Academy of Sciences, 48 Pyatnitskaya St., 119017, Moscow, Russia}

\date{Accepted 2017 March 28. Received 2017 March 26; in original form 2017 February 6}
\pubyear{2017}

\begin{document}
\label{firstpage}
\pagerange{\pageref{firstpage}--\pageref{lastpage}}
\maketitle

\begin{abstract}
In this paper, we extend the study initiated in Paper~I by modelling grain ensemble evolution in a dynamical model of an expanding \hii region and checking the effects of momentum transfer from dust to gas. The radiation pressure on the dust, the dust drift, and the lug on the gas by the dust are all important process that should be considered simultaneously to describe the dynamics of \hii regions. With accounting for the momentum transfer from the dust to the gas, the expansion time of the \hii region is notably reduced (for our model of RCW~120, the time to reach the observed radius of the \hii region is reduced by a factor of 1.5). Under the common approximation of frozen dust, where there is no relative drift between the dust and gas, the radiation pressure from the ionizing star drives the formation of the very deep gas cavity near the star. Such a cavity is much less pronounced when the dust drift is taken into account. The dust drift leads to the two-peak morphology of the dust density distribution and significantly reduces the dust-to-gas ratio in the ionized region (by a factor of 2 to 10). The dust-to-gas ratio is larger for higher temperatures of the ionizing star since the dust grains have a larger electric charge and are more strongly coupled to the gas.
\end{abstract}


\vspace{-0.5cm}
\section{Introduction}

{\it IRAS} and {\it MSX} imaging has revealed that Galactic infrared (IR) emission often forms arc-like or ring-like structures \citep{vanburen,CohenGreen2001}. A wealth of such structures has been subsequently discovered with the {\em Spitzer} and {\it WISE} space observatories \citep{churchwell_06,simpson_12}. Having related IR ring nebulae, or IR bubbles, as they are being called now, to the 20-cm continuum radio emission, \cite{deharveng_10} found that most of them (86 per cent) apparently surround \hii regions, ionized by one or several OB stars.

A specific feature of the IR bubbles is the different morphology at the different IR wavelengths. Emission from a typical bubble at {\em Spitzer} 8\,$\mu$m band appears as a narrow and more or less well-defined ring with quite a sharp inner boundary, bordering an \hii region. This ring is also visible at 24\,$\mu$m and {\em Herschel} wavelengths, from 70\,$\mu$m and longer. On the other hand, the emission at the {\em Spitzer} 24\,$\mu$m band (as well as at WISE 12\,$\mu$m and 22\,$\mu$m bands) also has a significant contribution from the interior of the \hii region \citep{deharveng_10}. The details of the morphology of this inner 24\,$\mu$m emission differ somewhat from one region to another. Sometimes it looks like a central peak and sometimes it resembles a complete or partial ring, enclosed within the larger outer 8\,$\mu$m ring. While inner emission appears also in the far-IR, the dichotomy of the inner and outer emission is most prominent at 8 and 24 micron.

A number of explanations has been put forward for this dual (outer 8\,$\mu$m versus inner 24\,$\mu$m) IR morphology. The absence of 8\,$\mu$m emission from the \hii region is readily explained by the absence of the relevant carriers. The origin of the emission in the {\em Spitzer} 8\,$\mu$m band is widely believed to be related to polycyclic aromatic hydrocarbons (PAH) or to some other small particles with an aromatic substructure \citep[see e.g.][]{KwokZhang,Jones2013}. It seems natural to assume that ultraviolet (UV) radiation of a central star or stars completely destroys PAHs within the \hii region.

Effective photodissociation of PAHs by UV radiation is supported both by theoretical estimates \citep[e.g.][]{Allain1996} and by observations \citep{Giard1994}. This is why bright PAH emission is only visible from a shell at the border of the \hii region. Closer to the star small, aromatic particles should be photodissociated, while further from the star, the radiation field is not strong enough to excite IR transitions. This assertion has been studied numerically by \cite{PavlyuchenkovAR} with a hydrodynamic model of an expanding \hii region and appropriate radiation transfer post-processing. It has been shown that the model does severely over-predict 8\,$\mu$m emission from within the \hii region under the assumption that dust is frozen to gas. This implies that PAHs should be absent from the \hii region somehow. In particular, an outer ring of 8\,$\mu$m emission can be reproduced by a model with a PAH destruction time-scale of the order of a few times $10^7$\,Myr for the general interstellar radiation field \citep{mmp83} (scaled appropriately for the radiation field inside the \hii region).  

The nature of the inner mid-IR emission (both at 12--24\,$\mu$m and at longer wavelengths) is more obscure. It may be related either to equilibrium heating of big grains or to stochastic heating of small grains. The model from the work of \cite{PavlyuchenkovAR} reproduced the intensity of the inner mid-IR emission, but provided no explanation to why the mid-IR emission should form not a central spike, but rather a ring (or an arc) {\em within\/} the \hii region.

As illumination conditions should vary smoothly in the ionized region, the specific emission morphology may reflect the distribution of the dust itself. \cite{dustwave} proposed that in some \hii regions arc-like structures are formed by dust, which is carried along by ionized gas, flowing towards the opening in the shell of the region, and is pushed away from the star by its radiation pressure. While such dust waves may lie at the origin of IR arcs in some regions, they do not explain complete 24\,$\mu$m rings like the one observed in the region N49, described, for example, in the work of \cite{watson_08} (see their fig.~7).

One of the popular explanations of the inner rings and arcs at 24\,$\mu$m relates their origin to the stellar wind from  an ionizing star (or a group of stars). Recently, \cite{Mackey_16} presented a set of 1D and 2D radiation hydrodynamics simulations, that reproduce the observed morphology of the mid- and far-IR emission towards RCW~120 under the assumption that the stellar wind produces the inner cavity. While the stellar wind can indeed be responsible for inner cavities in some \hii regions, the parameters of the wind are not well constrained. Also, to confirm this mechanism, direct identification of stellar winds is highly desirable, e.g. by observing the X-ray emission.

In \cite{paper_mnrasi} (hereinafter Paper~I) we considered the mechanism, that might have caused dust grains of different sizes to be distributed unevenly within the \hii region. This mechanism is the charged dust drift through ionized gas under the combined action of radiation pressure from the central star and gas drag (including Coulomb drag). In \citetalias{paper_mnrasi}, we showed that big grains ($a\approx3000$\,\AA) are effectively swept out of the \hii region by the radiation pressure. PAHs and very small grains ($a\approx30$\,\AA) are mostly coupled to the gas. If the UV radiation is low enough to allow near-zero PAH charge, PAHs can be removed from the \hii region during their neutrality periods, caused by charge fluctuations. For very small grains, the effect of charge fluctuations is less pronounced. The most interesting case is represented by intermediate-sized grains ($a\approx200$\,\AA). They accumulate at two stagnation radii producing a double ring structure with a less prominent inner ring and a denser outer ring.

It is tempting to relate these stagnation radii to the observed two-ring morphology of IR bubbles. However, this supposed relation cannot be substantiated without a proper modelling of the dust ensemble. In \citetalias{paper_mnrasi}, each grain was treated as if it were the only grain in the region. To compute the dust density distribution, one has to be more specific about the total number of grains of each kind included in the model, so here we adopt a much more detailed dust size distribution. Also, the necessity of a more accurate dust motion modelling calls for a refinement of the dynamical model. Specifically, one needs to account accuratly for momentum transfer from grains to gas via gas drag. In \citetalias{paper_mnrasi}, the dust ensemble did not influence the gas dynamics. 

In this paper, we extend the study initiated in \citetalias{paper_mnrasi} by modelling grain ensemble evolution in a dynamical model of an expanding \hii region and checking the effects of momentum transfer from dust to gas.

\vspace{-0.5cm}
\section{Model description}\label{sec_model}

The MARION model of an expanding \hii region is based on the Zeus-2D code \citep{stone_92}. It has been presented in detail in \citet{paper_hiimodel} and in \citetalias{paper_mnrasi}. Here we describe only the upgrades of the model relevant for the current study.

\subsection{Refined dust model}

\begin{figure}
\centering
\includegraphics[width=0.45\textwidth]{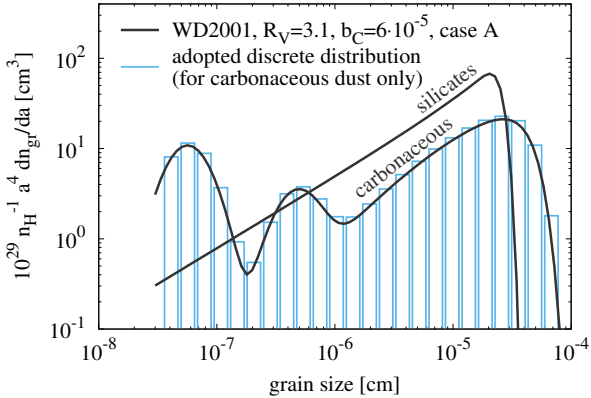}
\caption{Adopted initial silicate and carbonaceous grain size distributions and the discrete representation for carbonaceous grains (the discrete distribution for silicates is not shown).}
\label{fig:discretedust}
\end{figure}

In \citetalias{paper_mnrasi}, we considered four dust components, which are basically three conventional components (PAHs, 5\,\AA, very small carbonaceous grains, 30\,\AA, and big silicate grains, 3000\,\AA) and an intermediate-sized carbonaceous component with $a=200$\,\AA. This intermediate-sized component differs quite appreciably from the conventional components in its interaction with radiation and gas, so the more detailed consideration of the entire size range is desirable. Also, the limited number of components in the model can potentially be insufficient to fit the observed intensity maps. Thus, we added a more detailed dust size distribution in the model, with the whole size range for both silicate and carbonaceous grains divided into $N$ bins, so that the entire dust ensemble is represented by $2N$ components. We adopt $N=24$ for the current study. Size limits are from 4.2\,\AA\ to 0.9\,$\mu$m for carbonaceous grains and from 12\,\AA\ to 0.9\,$\mu$m for silicate grains. Carbonaceous grains are further subdivided into PAHs (size less than 50\,\AA) and graphite grains (the material defines the optical properties used to calculate the radiation pressure force and charging). We adopt grain material densities equal to 2.24~g/cm$^3$ for carbonaceous grains and 3.5~g/cm$^3$ for silicate grains. The amount of dust in each bin is scaled according to a model by \cite{WD2011} with parameters $b_{\rm C}=6\cdot10^{-5}$ and $R_{\rm V}=3.1$ at the beginning of the \hii region expansion. The initial dust size distributions adopted for carbonaceous and silicate grains are shown in Fig.~\ref{fig:discretedust}. As the region expands, the actual distribution may vary at each spatial location due to differential dust drift.

\subsection{Dust dynamical influence to gas}

Unless otherwise noted, we assume that dust grains are pushed away from the star by radiation pressure and slowed down by gas collisional and Coulomb drag. To compute the Coulomb drag, we use time-dependent electron and ion densities, which come from the detailed ionization balance included in MARION. To calculate the dust charge evolution, we use a refined model from \citetalias{paper_mnrasi} (see details in Appendix A). Grain charge is assumed to fluctuate around a mean value that is determined by the physical conditions at the current location of a specific grain.

In \citetalias{paper_mnrasi}, the gas dynamics was assumed to be unaffected by the dust dynamics. In the presented model, we account for the momentum transfer from dust to gas by including the corresponding drag force in the equation of gas motion (see Eq.~(3) from \citetalias{paper_mnrasi}). The momentum transfer from dust to gas is sensitive to the dust-to-gas ratio~$f_{\rm d}$. For the adopted dust model, the initial dust-to-gas ratio is $f_{\rm d}=f_{\rm d}^{\rm C}+f_{\rm d}^{\rm Si}=0.0024+0.0064$ and changes during the expansion of the \hii region.

\subsection{A set of models}

As in \citetalias{paper_mnrasi}, we assume that an ionizing star is initially embedded in a purely molecular medium with a uniform density of $3\cdot10^3$~cm$^{-3}$. We consider models with a main sequence star having an effective temperature $T_{\rm eff}$ of 30\,000\,K, 35\,000\,K, and 40\,000\,K. Stellar spectra are taken from~\citet{kurucz_79}. We assume that the spectrum of the star does not change during the expansion of the \hii region. Rough estimates of time-scales for hydrogen burning in stars with $T_{\rm eff}$ of 30\,000\,K, 35\,000\,K and 40\,000\,K are 4, 2 and $0.4 \cdot 10^6$\,years respectively \citep[see][Eq. 30.2 and Fig. 30.6]{Kippenhahn}. These time-scales are longer than the characteristic times we encounter in this study, so our assumption that the spectrum of the central star does not change appreciably during the \hii region expansion and corresponds to the main sequence seems to be realistic. While the stellar spectrum varies during the main sequence phase \citep[see e.g. recent results of][]{MartinsPalacios2017}, this effect is still not important for the time-scales in our present calculations. The change of the spectrum is an interesting issue that may produce some systematic effects or irregularities in the dust distribution, but it deserves a separate study.

The model with $T_{\rm eff}=35\,000$\,K is considered as the primary model. We describe three varieties of this model. In the frozen dust model, we simulate a thermal expansion of the \hii region aided by the radiation pressure on the dust under the assumption that dust is frozen to the gas. In this model, there is no relative drift between the dust and the gas, so that the radiation pressure effectively acts directly on the gas. In the coupled dust model, a relative dust-gas drift is accounted for. Also, we consider the no-dust-impact model in which there is no momentum transfer from the dust to the gas and the only factor driving the \hii region evolution is thermal expansion. This is the model presented in \citetalias{paper_mnrasi}. The models with their respective designations and some parameters is given in Table~\ref{models}. We intentionally neglect dust destruction for clarity and concentrate on pure dust drift.

\begin{table*}
\caption{Input and output parameters of models considered. Data shown in columns 5 and 6 correspond to the moment when the \hii region reaches 1.4~pc in size. The dust-to-gas ratio in column 5 is averaged over the volume within 1.4~pc. The inner gas cavity depth is calculated as the ratio between the \hii density at 1.4~pc and at 0.01~pc.}
\begin{tabular}{l|l|l|l|l|l}
\hline
Model  & $T_{\rm eff}$, K & Included processes & Time to reach 1.4~pc & Dust-to-gas ratio$, \%$ & Gas cavity depth\\
\hline
no dust impact               & 35\,000 & drift                   & 630 kyr &  0.17               &   1 (no cavity)     \\
frozen dust                  & 35\,000 & momentum transfer       & 400 kyr &  0.88 ($=$ initial) &   $\gg$10   \\
coupled dust                 & 35\,000 & drift+momentum transfer & 410 kyr &  0.26               &   2 \\
coupled dust [30]            & 30\,000 & drift+momentum transfer & 610 kyr &  0.11               &   1   (no cavity)   \\
coupled dust [40]            & 40\,000 & drift+momentum transfer & 265 kyr &  0.43               &   10  \\
\hline
\end{tabular}\\
\label{models}
\end{table*}

As in \citetalias{paper_mnrasi}, we present results for the physical time when the \hii region reaches the radius of about 1.4~pc, that is, approximately the radius of RCW~120. The radius is somewhat arbitrary, and more so, RCW~120 may not even be a good choice of a standard object to be compared with the simulation results (see Discussion). However, in this study, we are mostly interested in typical outcomes, and we postpone an analysis of real objects to the next paper.

\vspace{-0.5cm}
\section{Results}

\begin{figure}
\centering
\includegraphics[width=0.45\textwidth]{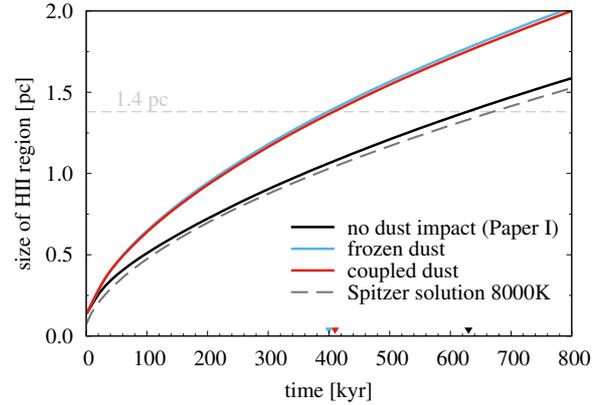}
\caption{The radius of the ionized region as a function of time for the different varieties of the model with $T_{\rm eff}=35\,000$\,K.}
\label{radiushiiregiontime}
\end{figure}

In our model, the expansion of the \hii region is driven by the combination of thermal and radiation pressure. The pure thermal expansion is a well-studied process in astrophysics, and there are a number of analytic expressions that describe it. In  Fig.~\ref{radiushiiregiontime}, we relate the radius of the ionized region as a function of time, $R(t)$, in the no-dust-impact model with that in the so-called Spitzer solution \citep{spitzerbook} for a gas temperature in the ionized region of $T_{\rm gas}=8000$~K. The Spitzer solution is very close to the numerical solution from the model with no impact from dust to gas. Given that the Spitzer solution does not account for the radiation pressure, we consider this as supporting the validity of our dynamical model.

To estimate the effect of radiation pressure on the dynamics of the \hii region in Fig.~\ref{radiushiiregiontime}, we also show $R(t)$ for models with frozen and coupled dust. The expansion rates are nearly the same in these two models, which is expected since in both cases nearly the same momentum is transferred from the dust to the gas. The model with no dust impact expands significantly more slowly, so that the time needed to reach a radius of 1.4~pc (indicated with a gray dashed line in the figure) is about 1.5 times greater in the no-dust-impact model than in the frozen dust and coupled dust models. That indicates the dynamical importance of the  radiation pressure.

\begin{figure*}
\centering
\includegraphics[width=0.95\textwidth]{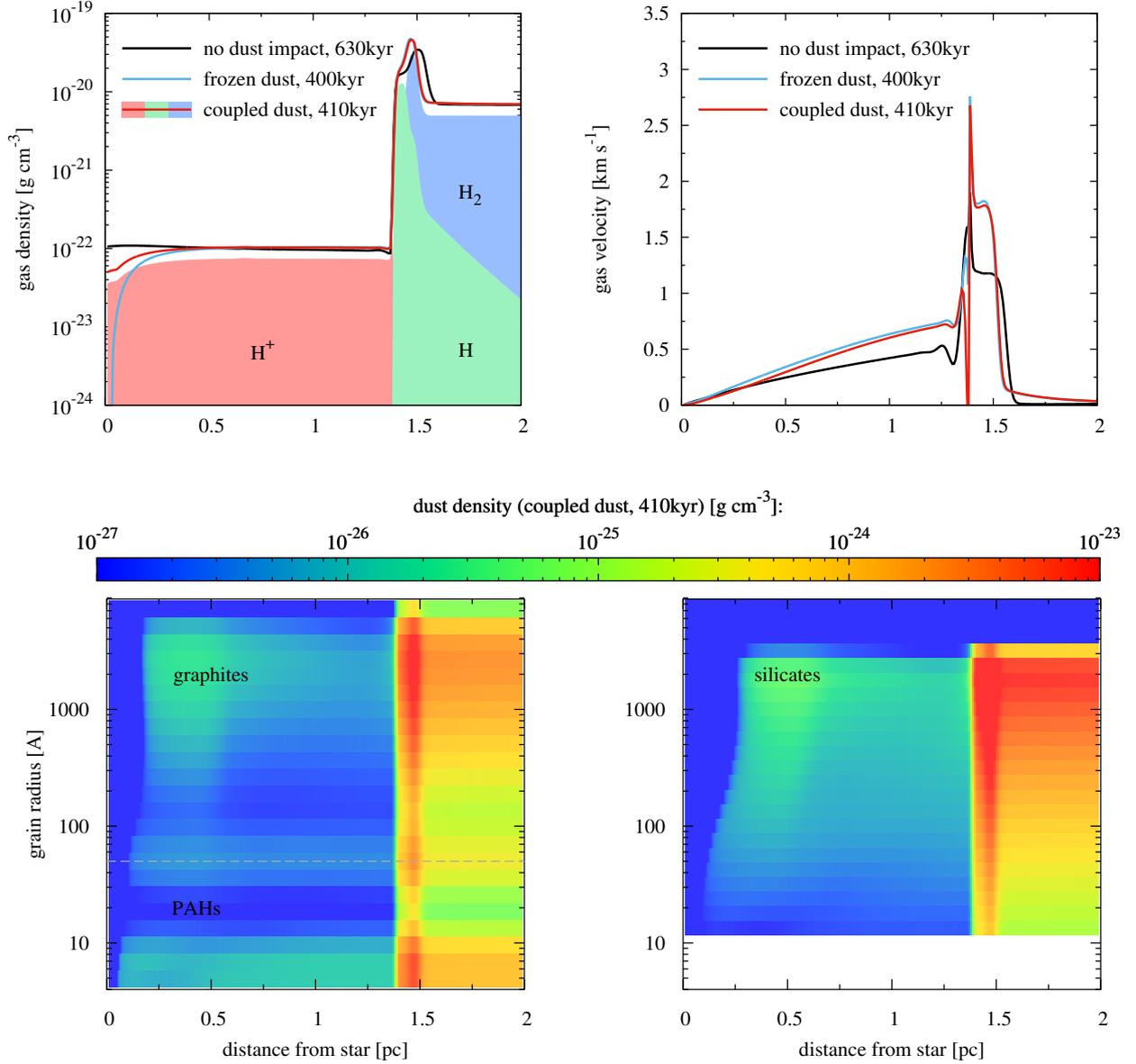}
\caption{Top row: Summary of the physical structure of the modelled region, when the \hii region radius is approximately equal to 1.4~pc, for the models with a stellar temperature of 35\,000\,K and different treatments of dust-gas coupling. Left: distributions of the total density (coloured lines) as well as of ionized, neutral, and molecular hydrogen (shaded areas, for the coupled dust model only) are shown. Right: velocity distributions. Bottom row: Diagrams showing carbonaceous (left) and silicate (right) dust size distributions as a function of radius at the final time moment for the coupled dust model. The size distributions at the right borders of the diagrams are almost identical to the adopted initial distributions.}
\label{genphys}
\end{figure*}

A general physical structure of the modelled \hii region and surrounding gas at the final time moment for the models with the star temperature $T_{\rm eff}=35\,000$\,K is shown in the top row of Fig.~\ref{genphys}. Shaded colour areas show distributions of various hydrogen components for the coupled dust model. An ionized gas region is surrounded by a thin envelope of atomic hydrogen. A dense shell, which has been swept by the shock, preceds the ionization front. It consists partially of atomic and partially of molecular hydrogen. The density contrast between the dense envelope and undisturbed molecular gas is $\approx7$.

If dust does not entrain the gas in motion (the no-dust-impact model, analogous to the one considered in \citetalias{paper_mnrasi}), the distribution of gas inside the \hii region is nearly flat (black line in Fig.~\ref{genphys}, top left) at the final time moment and, actually, throughout the expansion of the region as the gas dynamics is not affected by the dust motion in this case.

In contrast, for the model where the dust and the gas are tightly coupled ('frozen dust'), the dust under the action of radiation pressure transfers its momentum to the gas, which results in efficient removal of both the gas and the dust from the innermost part of the \hii region (the gas density for this model is shown with a blue line in Fig.~\ref{genphys}, top left). As a result, the immediate vicinity of the star is nearly devoid of matter. The size of this cavity grows very slowly, staying below $\sim0.1$~pc for the entire computational time. It takes about 400 kyr for the shell to reach the radius of 1.4~pc, that is, much less than in the no-dust-impact model, as dust pushes the gas, forcing it to expand faster.

In the coupled dust model, the dust transfers the same momentum to the gas (under our approximation of terminal velocity), but an inner cavity in the gas distribution is like the one in the frozen dust model at the earliest phase of expansion ($t\la1$~kyr). Then, dust expulsion, which is efficient in the coupled dust model, removes the dust from the centre of the region. In other words, the gas slips through the dust (in the reference frame, moving with dust) and stays within the inner dust cavity preserving a smoother gas density distribution. By the end of the computation, the depression of the inner gas density  is significantly less pronounced than in the model where the dust is assumed to be frozen to the gas.

As we have already mentioned, an important difference of the model with dust-gas coupling from the model presented in \citetalias{paper_mnrasi} is the significantly shorter evolutionary time-scale. While in the no-dust-impact model it takes 630~kyr for the dense shell to reach the RCW~120 radius, in the coupled dust model this time is 410~kyr only, that is, more than 1.5 times shorter. Accordingly, the maximum gas velocity associated with the layer of atomic hydrogen is about 2.7~km s$^{-1}$ in the coupled dust model and in the frozen dust model, being less than 2~km s$^{-1}$ in the no-dust-impact model (as shown in Fig.~\ref{genphys}, top right). The velocity of the molecular gas preceding the shock front is around 1.5~km s$^{-1}$ in the coupled dust and frozen dust model and is slightly above 1~km s$^{-1}$ in the no-dust-impact model.

In the bottom row of Fig.~\ref{genphys} we show the density distributions for all the dust components in the coupled dust model. All the grains are almost completely swept out from the centre of the \hii region (dust density is less than $10^{-27}$ g cm$^{-3}$ there). However, the size of the inner dust-free cavity is different for different grain types, being greater for larger grains, as they are more easily transported under the joint effect of radiation pressure and gas drag. Consequently, the size of the inner cavity is about 0.05~pc for smallest PAHs (i.e. 4 per cent of the total size of the \hii region) and more than 0.2~pc for big silicate grains with radii in excess of 0.2 $\mu$m (16 per cent of the total size of the \hii region). The size of the dust-free region is somewhat smaller for carbonaceous grains due to their better coupling to the gas. We adopted 4.4~eV for the work function of photoelectrons for carbonaceous grains and 8~eV for silicate grains~\citep{2001ApJS..134..263W}, so carbonaceous grains are more prone to the photoelectric charging and, hence, better coupled to the ionized gas.

The significant depletion of both gas and dust in the frozen dust  model is quantitatively similar to the distribution obtained by \cite{Rodr}. It is easy to see that in the coupled dust model the central gas depletion zone is much shallower, while dust is swept away from a somewhat large volume. The notable feature of the coupled dust  which is not observed in the frozen dust model is the enhancement of dust density over 0.2--0.6 pc (light-green areas in the bottom panels of Fig.~\ref{genphys}) with respect to the nearly flat density distribution over 0.6--1.4 pc. A similar inner density ring is formed in the no-dust-impact model, discussed in detail in \citetalias{paper_mnrasi}. We argue that this kind of morphology may be partially responsible for the formation of inner emission rings at 24~$\mu$m seen toward some \hii regions, but this is a problem for a separate study.

\begin{figure}
\centering
\includegraphics[width=0.45\textwidth]{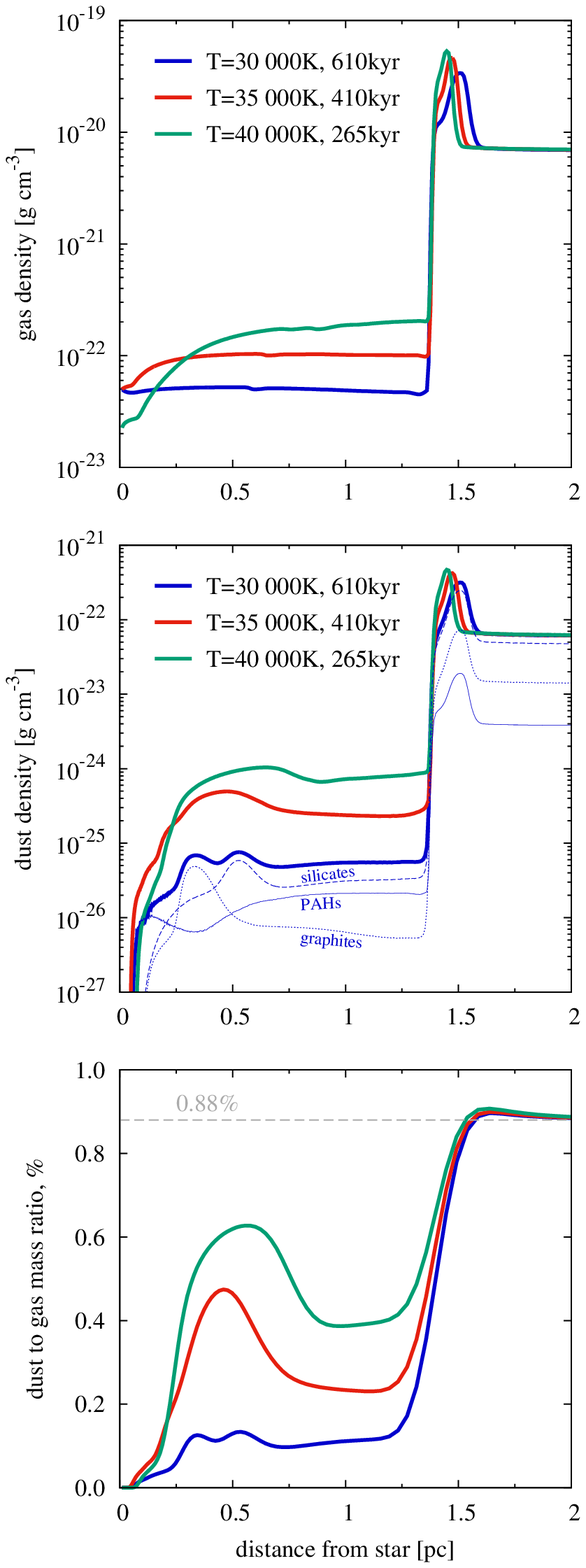}
\caption{Radial distributions of gas density (top), dust density (middle), and dust-to-gas mass ratio (bottom) in models with $T_{\rm eff}=30\,000$\,K (blue), 35\,000\,K (coupled dust model, red), and 40\,000\,K (green). The gray dashed line in the bottom plot indicates dust-to-gas ratio in the frozen dust model (identical to the initial value).}
\label{dust-gas-stars}
\end{figure}

The sequence of models no-dust-impact $\rightarrow$ frozen dust $\rightarrow$ coupled dust represents a gradual improvement in accounting for the dynamical effect of the dust on the gas. While the first advance significantly affects the expansion time-scale and density structure, the second advance is important for the density structure in the very vicinity of the star. In the following, we utilize the most realistic model -- the coupled dust model -- to investigate the dependence of the structure of the \hii region on the central star energetics. We consider two models which differ from the coupled dust model by $T_{\rm eff}$ value of 30\,000\,K and 40\,000\,K. The gas and dust density distributions for these models are compared to the coupled dust model in Fig.~\ref{dust-gas-stars}. The gas density is nearly constant inside the \hii region for the lowest $T_{\rm eff}$. The shallow central depression in the gas distribution, which still exists in the model with a star temperature of $T_{\rm eff}=35\,000$\,K by the end of the computational time, appears in the coupled dust model only at the earliest phase and vanishes completely after $10^5$ yr of evolution. Due to the weaker radiation pressure, it takes about 610~kyr for the shell to reach the radius of 1.4~pc.

The effect of radiation pressure and dust-gas coupling is most pronounced in the model with $T_{\rm eff}=40\,000$\,K. In this case, the gas density drops significantly in the vicinity of the star, as the gas is dragged along with dust, while the expansion time is 265~kyr only. If we continue the computation further, we find that the central gas depression smooths out as the expansion proceeds, but the region of lower gas density around the star is still visible even after 1 Myr of evolution. Beyond the immediate vicinity of the star, the gas density inside the \hii region is highest in the model with the hottest star as gas swept away from the cavity stays within the \hii region.

Dust is almost completely swept out of the centre of the \hii region in all coupled dust models, resulting in inner holes in the dust distributions. Beyond the central cavities, the dust distribution has an inner bump of moderate height inside the \hii region (at $\approx0.5$\,pc for the considered cases and times), a flat plateau (from $\approx0.7$\,pc up to the border of the ionized region at $1.4$\,pc) and a strong bump in the area of the collected neutral gas ($1.4-1.6$\,pc). Due to the closeness of the hot central star, the dust temperature inside the inner lower-density dust bump is higher than in the outer high-density bump. Thus, the IR flux from the inner bump can be comparable to the emission from the outer bump. This would result in a double-ring intensity distribution. The effect of such two-peak dust morphology (a low density bump at 0.5\,pc and a high density bump at 1.5\,pc) on the IR images of \hii regions deserves a separate study. Note that the inner bump can also have substructure, as seen in the model with $T_{\rm eff}=30\,000$\,K due to the contribution from silicate and graphite dust populations. However, the action  of other neglected processes (e.g. dust destruction, stellar wind, 3D morphology) may suppress such substructures. 

Somewhat counter-intuitively, the dust density within the \hii region is highest in the model with highest $T_{\rm eff}$ and lowest for the star with $T_{\rm eff} =30\,000$\,K. One might expect that stronger radiation would expel dust most effectively, but another effect wins in this case. Specifically, the dust positive charge is higher in the model with the hotter star, cf. Fig. 1 in \citetalias{paper_mnrasi}. Charged dust is more tightly bound to the plasma in the \hii region in the model with $T_{\rm eff} =40\,000$\,K, and so it is more prone to the Coulomb drag force.

In the middle panel of Fig.~\ref{dust-gas-stars}, we show individual contributions from various dust components for the model with $T_{\rm eff}=30\,000$\,K. The enhancement of the dust density consists of three peaks in this case, corresponding to PAHs, graphites and silicates. Due to the less violent radiation pressure in this model, spatial distributions of various dust grains are sensitive to subtle details of their interaction with the plasma and radiation.

Dramatic changes of the dust-to-gas mass ratio $f_{\rm d}$ inside the \hii region are shown in the bottom panel of Fig.~\ref{dust-gas-stars}. The ratio in the undisturbed molecular region is close to the initial value of 0.88 per cent value. There is no significant dust drift in the neutral region. The increase of $f_{\rm d}$ at $\approx1.6\,$pc over the initial value represents the dust swept from the \hii region. This bump is quite small due to the 3--5 orders of magnitude difference in the dust densities inside and outside the \hii region. So, the blown-out dust represents only a small addition to the dust shoveled by the shock wave. Dust-to-gas mass ratios within the ionized region are, in contrast, significantly lower than the initial value. Within 0.2~pc from the star, the dust-to-gas ratio drops below 10$^{-3}$ in all three coupled dust models. At $0.2<r<1.4$~pc, the smallest values of the dust-to-gas ratio are observed in the model with the lowest $T_{\rm eff}$.

The average dust-to-gas mass ratios inside the \hii region are 0.11, 0.26 and 0.43 per cent for the coupled dust models with $T_{\rm eff}$ equal to 30\,000, 35\,000 and 40\,000~K, respectively. These values, corresponding to when the size of the \hii region reaches 1.4~pc, are summarized in Table~\ref{models}. The dust-to-gas ratio for the no-dust-impact model (Paper I) is 0.17\,per\,cent, which is lower than 0.26\,per\,cent value in the more realistic coupled dust 35\,000~K model and stresses the importance of dust-gas interaction neglected in Paper I. 

\begin{figure}
\centering
\includegraphics[width=0.45\textwidth]{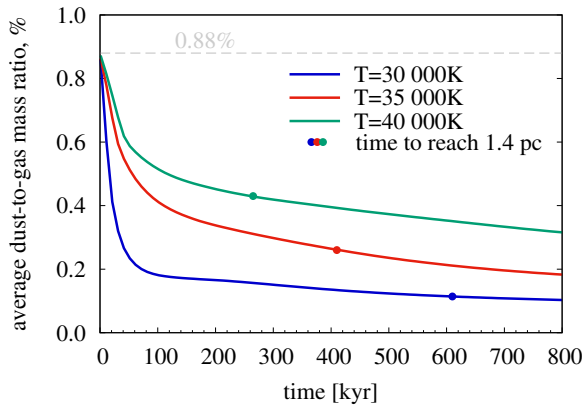}
\caption{The evolution of the average dust-to-gas mass ratio inside the ionized region in models with $T_{\rm eff}=30\,000$\,K (blue), 35\,000\,K (red), and 40\,000\,K (green). Dots represent time when the \hii region reaches a size of 1.4~pc.}
\label{fig5}
\end{figure}

The evolution of the average dust-to-gas mass ratio inside the ionized region for various values of $T_{\rm eff}$ is presented at Fig.~\ref{fig5}. The time-scales for the blow-out of half of the dust mass are $20, 100$ and $250$\,kyr for the models with $T_{\rm eff}=30\,000, 35\,000$ and $40\,000$\,K, respectively. While the dust is significantly swept away from the \hii region into the neutral shoveled envelope, the dust drift in the molecular region itself is inefficient, and the dust-to-gas mass ratio there is not affected by the radiation pressure. The relative dust/gas velocity for the most mobile sub-micron grains is less than 0.03 km s$^{-1}$ in the molecular region. However, we need to keep in mind that we consider an \hii region that expands into a uniform medium. The situation would be different if the expansion occured in a medium with a declining or tapered density distribution (e.g. close to the border of a molecular cloud). In this case, we may expect a less straightforward evolution, including gas-dust segregation, leading to observable dust/gas variations in the interstellar medium.

\vspace{-0.5cm}
\section{Discussion}

In this paper, we consider how radiation pressure changes the dynamical and physical structure of an expanding \hii region near a single O or B-star. The key feature of our model is that we account for the radiation pressure on dust and for the momentum transfer from dust to gas. We do not consider the direct influence of the radiation pressure on gas. \cite{Krumholz_09} and \cite{Draine2011} showed that radiation pressure on the gas gives an unimportant contribution to the dynamics of an \hii region excited by a star comparable to the one ionizing RCW~120 or other IR bubbles. The spectral type of the ionizing star in RCW~120 (CD-38,11636) has been determined several times in the past. For example, \citet{georgelin_70} and \citet{Avedisova_84} classified it as O8V, \citet{Martins_10} found an earlier spectral type of O6-8V/III and derived $T_{\rm eff}=(37.5 \pm 2)\times 10^3$\,K. In this study we adopt $T_{\rm eff}=35000$\,K for most our models, which corresponds to the rate of emission of ionizing photons $Q_{0}= 0.13 \cdot 10^{49}$~s$^{-1}$ \citep{1984ApJ...283..165T}. According to \citet{Draine2011}, the impact of the  radiation pressure on the gas is only significant for $Q_{0,49}\cdot n\ga10^3$, where $Q_{0,49} = Q_0 /10^{49}$. The typical gas number density $n$ inside the \hii region in our calculation does not exceed 200~cm$^{-3}$, so that $Q_{0,49}\cdot n<26$~cm$^{-3}$, which means that the radiation pressure on the gas is insignificant in our modeling. 

\citet{Kim_2016} simulated the time-dependent expansion of \hii regions, taking into account the radiation pressure along with the gravitational slow-down of the expansion. The physical definition of their model corresponds to our frozen dust model. They also found that for $Q_{0,49}\cdot n < 10^4$, the radiation pressure on the gas does not have a significant impact on the gas density distribution inside the \hii region. \citet{Kim_2016} noted that the effect of the radiation pressure leads to compression of the ionized gas to the border of the \hii region, which looks like a rise in the gas density at the outer part of the \hii region. We see a similar structure in our calculations. Dust does not escape the \hii region but  rather is redistributed inside it. However, our modeling shows that in an accurate calculation of dust and gas coupling, the formation of a deep inner cavity in the gas distribution around a star with $T_{\rm eff}=30\,000 - 35\,000$\,K is prevented, in contrast to the results of~\citet{Kim_2016}. When dust escapes the immediate vicinity of the star, the radiation pressure does not affect the matter in this area, and ionized gas reappears in the centre of the region.

The expansion time for the \hii region is smaller  by about a factor of 1.5 relative to our previous calculation without radiation pressure~\citepalias{paper_mnrasi}. While in the original model of~\citetalias{paper_mnrasi} the dynamical age estimate for RCW~120 was 630~kyr, it is only 410 kyr in the coupled dust model considered here. RCW~120 was proposed as a region where triggering of star formation via the collect-and-collapse mechanism has taken place in the past \citep[e.g.][]{zavagno_07}. For this mechanism actually to have been responsible for triggering star formation in RCW~120, young stellar objects should have been formed about 300~kyr.

Our results imply that one should use with caution the visible size of an \hii region to estimate   its dynamical age using analytic solutions like the Spitzer solution. Also, conclusions about the triggering via the collect-and-collapse mechanism should take into account accurate calculation of radiation pressure. For example, \cite{Kirsanova_14} used the visible sizes of some \hii regions to infer the possibility of collect-and-collapse mechanism, using theoretical expressions from \citet{Whitworth_94}. They concluded that collect-and-collapse triggering is marginally possible in the S235 region only if the density of the surrounding medium is quite high ($\approx7\cdot10^3$~cm$^{-3}$). The expressions in \citet{Whitworth_94} are based on the Spitzer solution. If we used our solution instead, parameterized as $Kt^\alpha$, we would get a fragmentation time which is nearly a factor of 2 shorter, than the estimate in \cite{Kirsanova_14}, even for density $\approx3\cdot10^3$~cm$^{-3}$, making `collect-and-collapse' triggering more probable in this region.

Our final goal is to study whether the accurate calculation of gas and dust coupling can explain morphology of IR dust emission towards \hii regions. Here we would like to emphasize that radiation pressure can form a central dust-free volume inside an \hii region without an accompanying gas cavity. We are trying to follow the Occam's razor principle, so our present model is limited. It does not take into account stellar wind from the ionizing star. Moreover, IR bubbles on all scales are sometimes referred to as wind-blown bubbles \citep[see e.g.][]{churchwell_06}. We intend to show in our next paper that specific features of 24 $\mu$m emission from IR bubbles could be at least partly explained without the hypothesis of wind-blown cavities. It is important to note that the direct wind signature, that is X-ray emission, is only observed towards a very limited number of objects, e.g. in the Orion region by \citet{Gudel_08}.

\cite{2012ApJ...760..149P} reported the 24 $\mu$m peak emission towards MAGPIS 20~cm peak in a uniform sample of evolved \hii regions around single OB-stars with effective temperatures up to 40\,000 -- 45\,000\,K. They also expected to find the 24~$\mu$m emission peak between the 8~$\mu$m and 250~$\mu$m emission peaks because 24~$\mu$m emission is thought to be produced by dust grains with an intermediate size relatively to PAHs and big grains. They concluded that the peak of 24 $\mu$m emission is caused by a new generation of dust grains that were re-supplied by destruction of dense cloudlets embedded in the extended \hii regions. However, the presence of intermediate-sized grains inside the \hii region in our model does not require them th reappear. Motion of small charged graphite grains in \hii regions is not effective enough to make the ionized gas free from dust, at least around ionizing stars with $T_{\rm eff}$ down to 30\,000\,K.

\vspace{-0.5cm}
\section{Conclusion}
This paper is our follow-up theoretical study of \hii regions after \cite{paper_hiimodel}, \cite{PavlyuchenkovAR}, and \cite{paper_mnrasi}. We modified the model of the expanding \hii region by including the effect of momentum transfer from dust to gas and adopting the multicomponent model of dust. With this model we calculated the density, chemical and thermal evolution of the \hii region. Both the gas and dust dynamics are considered during the expansion of the \hii region. We account for charged dust drift through ionized gas under the combined action of radiation pressure from the central star and gas drag (including Coulomb drag). Our results can be summarized as follows:
\begin{enumerate}
\item The radiation pressure on the dust, the dust drift, and the lug on the gas by the dust are all important process that should be considered in describing the dynamics of \hii regions. By accounting for these processes, the expansion time of the \hii region is notably reduced (for our model of RCW~120, the time to reach observed radius of the \hii region is reduced by a factor of 1.5).

\item Under the frozen dust approximation (i.e. neglecting dust drift relative to the gas), the radiation pressure from the ionizing star drives the formation of a gas cavity near the star. When dust drift is taken into account the gas leaks through the dust and fills the cavity, resulting in a smoother gas density distribution.

\item Dust drift leads to a two-peak morphology of the dust density distribution with the inner low-density peak inside the \hii region and high-density peak at the outer boundary of the ionized region. While the radiation pressure expels 50--90\,per\,cent of the dust mass from the \hii region into the neutral shoveled envelope, the dust drift in the molecular region is inefficient. The time-scales of the dust blow-out are $20, 100$ and $250$\,kyr for the models with effective star temperatures of $T_{\rm eff}=30\,000, 35\,000$, and $40\,000$\,K, respectively. Notably, the dust-to-gas ratio is larger for an ionizing star with a higher temperature since the dust grains have a larger electric charge and are more strongly coupled to the gas.
\end{enumerate}

While the low dust-to-gas mass ratio inside the \hii region may be produced by dust (photo)destruction and by dust drift, these mechanisms depend on the temperature of the central star in a different way. A central star with a higher temperature should be more effective in dust photo-destruction, but less effective in removal dust by radiation pressure.

\vspace{-0.5cm} 
\section*{Acknowledgements}
We thank the referee for her/his helpful comments and fast revision. This work was supported by RFBR grants 16-02-00834 and 17-02-00521.

\vspace{-0.5cm} 
\bibliographystyle{mnras}
\bibliography{paper}

\vspace{-0.5cm} 

\appendix
\section{Grain charge} \label{AppA}

In \citetalias{paper_mnrasi}, to reduce computational time, the grain charge was pre-calculated on a grid of electron number densities, gas temperatures and distances from the star (or radiation field strength). To widen the versality of MARION, we incorporated the grain charge computational module in to the main code. The increased computational costs were circumvented by employing parallel algorithms.

\begin{figure}
\centering
\includegraphics[width=0.45\textwidth]{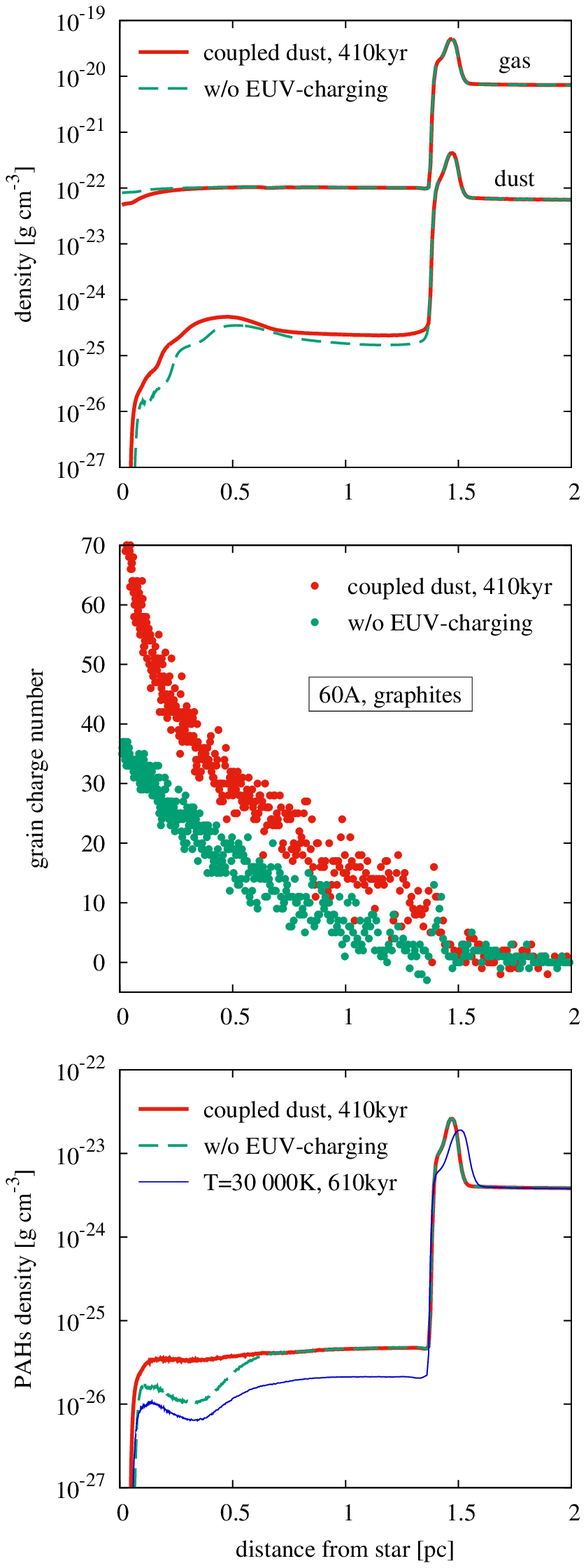}
\caption{Comparison between the fiducial coupled dust model (red) and the same model but without EUV photocharging (green). Upper: gas and total dust densities; Middle: charge of a $60$\,\AA~graphitic grain; Bottom: PAH density (see text).}
\label{FigA1}
\end{figure}

In the previous version of MARION, the EUV photons (extreme ultraviolet radiation, $h\nu>13.6$\,eV) were incorrectly neglected in the integration of the photocharging current. This led to an underestimation of the grain charge in the ionized region and did not have any impact in $\rm H$ and ${\rm H}_2$ regions due to the absence of EUV photons there. Here we demonstrate how fixing this issue affects the gas and dust dynamics. In the upper panel of Fig.~\ref{FigA1}, we show the gas and dust density profiles for the fiducial coupled dust model and for the same model but without accounting for EUV photons. In other words, the only difference between models is the upper integration limit in the Eq.~(A7) of \citetalias{paper_mnrasi}. The difference in the grain charge of $60$\,\AA~graphites (arbitrary choice) is shown in the middle panel of Fig.~\ref{FigA1}. The maximum difference is in the very inner part and reaches a factor of 2, while the average charges in the non-ionized regions are virtually the same (the spread is caused by the natural charge dispersion).

The higher grain charge in the ionized region leads to better dust retention, and, hence, slightly more efficient gas blow-out from vicinity of the star. While we assume that the improvement in the charge calculation  does not have a significant impact on the overall gas and dust densities, there is one qualitative difference. The PAHs do not leak through the ionized gas when their charges fluctuate near zero in the model with $T_{\rm eff}=35\,000$\,K (cee the red and dashed green lines in the bottom panel of Fig.~\ref{FigA1}). However, they still leak for a lower effective temperature of the star of 30 000K (lower photoionization current; see blue line in the bottom panel of Fig.~\ref{FigA1}). We note that there is also an inverse dependence of PAH retention in the \hii region on the assumed work function (we use $W=4.4$\,eV for both PAHs and graphites). So, we keep the conclusion from \citetalias{paper_mnrasi}, that there is a possibility to evacuating (some) PAHs from the \hii region due to near-zero charge fluctuation.

\bsp	
\label{lastpage}
\end{document}